\documentclass[aps,prl,preprint,showpacs,showkeys,superscriptaddress]{revtex4}
\usepackage{pstricks}
\usepackage{epsfig}
\usepackage{graphicx}
\usepackage{amsmath}
\usepackage{amsfonts}
\usepackage{amssymb}
\usepackage{rotating}
\newcommand{\bra}[1]{\langle #1|}
\newcommand{\ket}[1]{|#1\rangle}

\begin{document}
\title{Variational two-particle density matrix calculation for the Hubbard model below half filling using spin-adapted lifting conditions}
\author{Brecht Verstichel}
\email{brecht.verstichel@ugent.be}
\affiliation{Ghent University, Center for Molecular Modeling, Technologiepark 903, 9052 Zwijnaarde, Belgium}
\author{Helen van Aggelen}
\affiliation{Ghent University, Department of Inorganic and Physical Chemistry, Krijgslaan 281 (S3), B-9000 Gent, Belgium}
\author{Ward Poelmans}
\affiliation{Ghent University, Center for Molecular Modeling, Technologiepark 903, 9052 Zwijnaarde, Belgium}
\author{Dimitri Van Neck}
\affiliation{Ghent University, Center for Molecular Modeling, Technologiepark 903, 9052 Zwijnaarde, Belgium}
\begin{abstract}
The variational determination of the two-particle density matrix is an interesting, but not yet fully explored technique that allows to obtain ground-state properties of a quantum many-body system without reference to an $N$-particle wave function. The one-dimensional fermionic Hubbard model has been studied before with this method, using standard two- and three-index conditions on the density matrix [J. R. Hammond {\it et al.}, Phys. Rev. A 73, 062505 (2006)], while a more recent study explored so-called subsystem constraints [N. Shenvi {\it et al.}, Phys. Rev. Lett. 105, 213003 (2010)]. These studies reported good results even with only standard two-index conditions, but have always been limited to the half-filled lattice. In this Letter we establish the fact that the two-index approach fails for other fillings. In this case, a subset of three-index conditions is absolutely needed to describe the correct physics in the strong-repulsion limit. We show that applying lifting conditions [J.R. Hammond et al., Phys. Rev. A 71, 062503 (2005)] is the most economical way to achieve this, while still avoiding the computationally much heavier three-index conditions. A further extension to spin-adapted lifting conditions leads to increased accuracy in the intermediate repulsion regime.  At the same time we establish the feasibility of such studies to the more complicated phase diagram in two-dimensional Hubbard models.
\end{abstract}
\pacs{}
\keywords{}
\maketitle
%introduction, general many-body problem
The main problem in many-body quantum mechanics, which comprises nuclear physics, quantum chemistry and condensed matter physics, is the exponential increase of the dimension of Hilbert space with the number of particles. The challenge has therefore been to develop approximate methods which describe the relevant degrees of freedom in the system without an excessive computational cost, {\it i.e.} with a polynomial increase. In one of these methods, the $N$-particle wave function is replaced by the two-particle density matrix (2DM), and over the last decade, a lot of progress has been made in this field \cite{mazziotti,maz_prl,nakata_first,nakata_last,atomic,helen_2}. For a Hamiltonian:
\begin{equation}
\hat{H} = \sum_{\alpha\beta} t_{\alpha\beta}a^\dagger_\alpha a_\beta + \frac{1}{4}\sum_{\alpha\beta\gamma\delta}V_{\alpha\beta;\gamma\delta}a^\dagger_\alpha a^\dagger_\beta a_\delta a_\gamma~,
\end{equation}
containing only pairwise interactions, the energy of the system can be expressed as:
\begin{equation}
E(\Gamma) = \mathrm{Tr}~\Gamma H^{(2)} = \frac{1}{4}\sum_{\alpha\beta\gamma\delta}\Gamma_{\alpha\beta;\gamma\delta}H^{(2)}_{\alpha\beta;\gamma\delta}~,
\label{ener_func}
\end{equation}
in terms of the 2DM: 
\begin{equation}
\Gamma_{\alpha\beta;\gamma\delta} = \bra{\Psi^N}a^\dagger_\alpha a^\dagger_\beta a_\delta a_\gamma \ket{\Psi^N}~,
\label{2DM}
\end{equation}
and the reduced two-particle Hamiltonian,
\begin{equation}
H^{(2)}_{\alpha\beta;\gamma\delta} = \frac{1}{N-1}\left(\delta_{\alpha\gamma}t_{\beta\delta} - \delta_{\alpha\delta}t_{\beta\gamma} - \delta_{\beta\gamma}t_{\alpha\delta} + \delta_{\beta\delta}t_{\alpha\gamma}\right) + V_{\alpha\beta;\gamma\delta}~.
\end{equation}
Second-quantized notation is used where $a^\dagger_\alpha$ $(a_\alpha)$ creates (annihilates) a fermion in the single-particle state $\alpha$.

In variational density-matrix optimization (v2DM), originally introduced by L\"owdin, Mayer and Coleman \cite{lowdin,mayer,coleman}, one exploits this fact and uses the 2DM as a variable in a variational approach. From the resulting 2DM all one- and two-body properties of the ground state can be extracted. This should not be implemented naively, however, as there are a number of non-trivial constraints which a 2DM has to fulfil in order to be derivable from a $N$-particle wave function. This is the $N$-representability problem \cite{coleman} which was proven to belong, in general, to the QMA-complete complexity class \cite{qma}. In practical approaches one uses a set of conditions which are necessary but not sufficient, and therefore lead to a lower bound on the ground-state energy. The most commonly used are the two-index conditions, called $P$ (or $D$), $Q$ and $G$ \cite{coleman,garrod}, and the computationally much heavier three-index conditions called $T_1$ and $T_2$ \cite{zhao,hammond}. They all rely on the fact that for a manifestly positive Hamiltonian $\hat{H} = \sum_i\hat{B}_i^\dagger \hat{B}_i$, the expectation value of the energy has to be larger than zero. These conditions can be expressed as linear matrix maps of the 2DM that have to be positive semidefinite. Another type of constraint that has recently been developed are the subsystem or active-space constraints \cite{sebold,qsep,shenvi} in which linear conditions are imposed only that part of the density matrix that is related to a subspace of the complete single-particle space. This allows to increase accuracy (in the subspace) without having to use three-index conditions.
%hubbard, previous studies on half filled lattice, huge errors below half filling, first figure
Such v2DM methods have been used to study a wide variety of many-body systems: nuclei \cite{rosina}, atoms and molecules \cite{mazziotti,maz_prl,nakata_first,nakata_last,atomic,helen_2}, but also lattice systems \cite{kijewski,sebold,maz_hub,shenvi,nakata_last}. 

The Hubbard Hamiltonian \cite{hubbard} is the simplest schematic Hamiltonian that models the non-trivial correlations in solids as a competition between a delocalizing hopping term and an on-site repulsion term. In one dimension this Hamiltonian reads:
\begin{equation}
\hat{H}=-\sum_{a\sigma} \left(a^\dagger_{a;\sigma}a_{a+1;\sigma} + a^\dagger_{a+1;\sigma}a_{a;\sigma}\right) + U\sum_a a^\dagger_{a\uparrow}a_{a\uparrow}{a}^\dagger_{a\downarrow}a_{a\downarrow}~,
\label{hubbard}
\end{equation}
where the sites on a periodic lattice are labeled $a$ and $\sigma$ is the (up or down) spin.
In previous v2DM studies of the one-dimensional Hubbard model \cite{maz_hub,nakata_last,shenvi} only the half-filled lattice was studied, and it was found that even the two-index conditions could accurately describe the ground-state energy. In this Letter we show that the two-index conditions fail to describe the strong correlation limit below half filling, and that the subsystem constraints, as introduced in \cite{qsep,shenvi}, cannot solve this problem. In fact, a particular type of three-index conditions are needed in this limit, and we show that by using the 2.5DM (which is the 3DM diagonal in one spatial index) as the central object, these constraints can be incorporated while keeping the basic matrix manipulations in two-particle space.

%here the introduction is somewhat over, so now, figure of 6-5 hubbard, 
\begin{figure}
\centering
\includegraphics[scale=0.7]{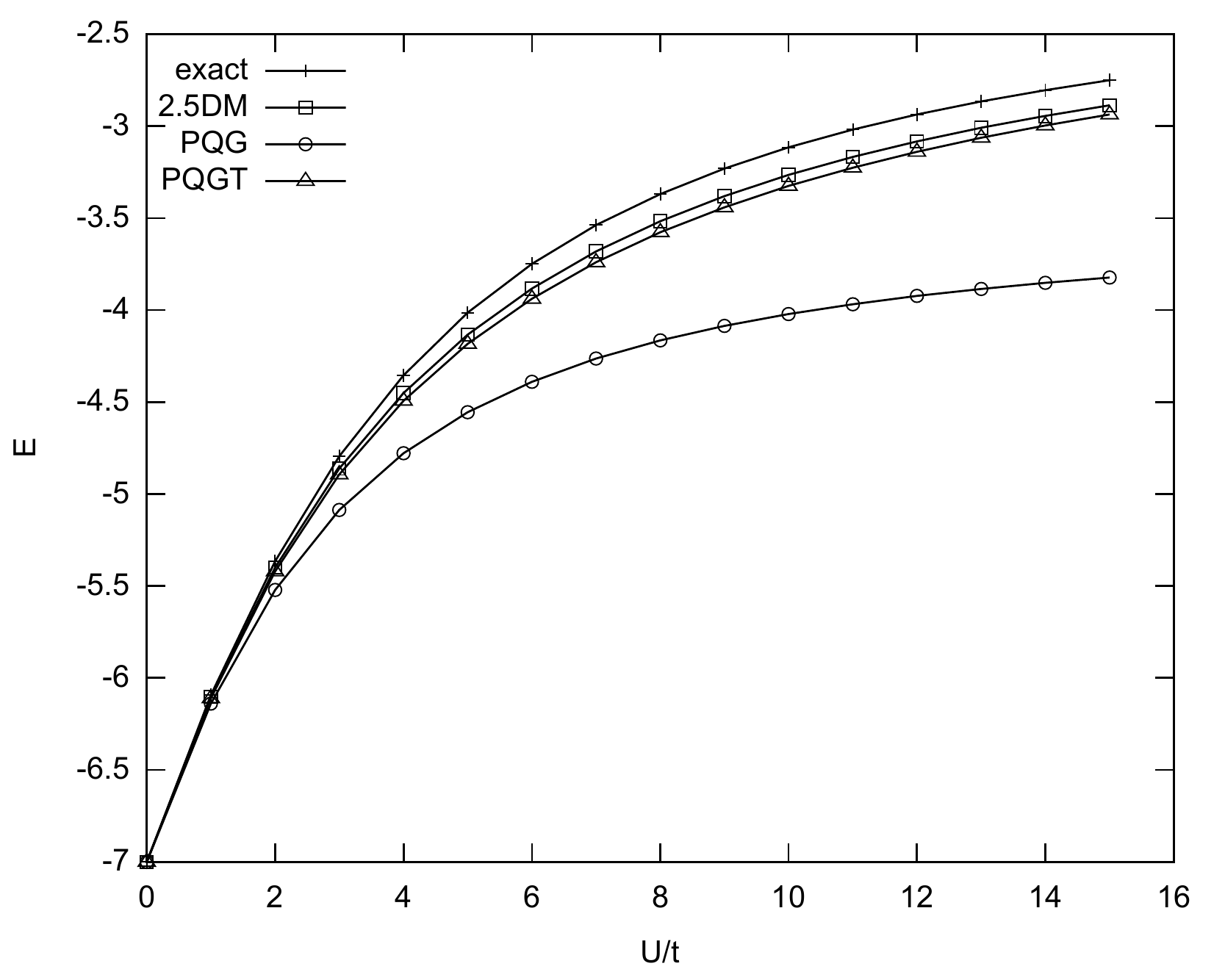}
\caption{\label{6_5} Ground-state energy as a function of on-site repulsion $U$ for 5 particles on a 6-site lattice. Exact results compared with v2DM results using $PQG$ and $PQGT_1T_2$, and the v2.5DM results.}
\end{figure}
\begin{table}
\centering
\begin{tabular}{c|cccc}
$U$ & $PQG$ & $PQGT$ &v2.5DM&exact\\
\hline
50 & -3.55 & -2.29 & -2.28 & -2.20\\
100 & -3.49 & -2.15 & -2.14& -2.08\\
1000 & -3.44 & -2.03 & -2.01& -2.01
\end{tabular}
\caption{ \label{tabel_5_6} Ground-state energy of a 6-site lattice with 5 particles for $U = 50,\ 100$ and 1000, exact results compared with v2DM results using $PQG$ and $PQGT$ results, and v2.5DM results.}
\end{table}
In order to demonstrate the problem, Fig.~\ref{6_5} shows the ground-state energy as a function of the on-site repulsion $U$, for 5 particles in a 6-site lattice. As one can see the general form of the exact $E$ vs. $U$ curve (obtained through diagonalization) is nicely described by the $PQGT$ result ({\it i.e.} two- and three-index conditions). On the other hand the $PQG$ result (only two-index conditions) grossly underestimates the energy when $U$ increases. The large-$U$ limit is examined in Table \ref{tabel_5_6} and one notes that $PQG$ fails to get the energy right in this limit, as opposed to the $PQGT$ result. When inspecting the $PQG$-optimized 2DM in the large-$U$ limit, it was found that the on-site repulsion term vanishes, as it should be, since the 2DM elements corresponding to doubly occupied sites are zero:
\begin{equation}
\lim_{U\rightarrow\infty}\Gamma^{PQG}_{a\uparrow a\downarrow;c\uparrow d\downarrow} = 0~.
\label{onsite_rep}
\end{equation}
So the problem with the two-index conditions lies in its inability to describe the hopping term on a lattice where the sites cannot be doubly occupied. It is readily understood that subsystem constraints cannot solve this issue, because the singly-occupied space is a subspace of the full $N$-particle Hilbert space, and cannot be obtained by a restriction of single-particle space to a subsystem. 

The creation and annihilation of particles on a singly occupied lattice can be described by the so-called Gutzwiller operators \cite{gutz_1,gutz_2}:
\begin{eqnarray}
g_{\alpha} &=& a_{\alpha}\left(1 - a^\dagger_{\bar{\alpha}}a_{\bar{\alpha}}\right),\\
g^\dagger_{\alpha} &=& \left(1 - a^\dagger_{\bar{\alpha}}a_{\bar{\alpha}}\right)a^\dagger_{\alpha}~,
\end{eqnarray}
where $\alpha$ and $\bar{\alpha}$ are single particle indices on the same site with opposite spin.
In analogy with the necessary and sufficient conditions for $N$-representability of the one-particle density matrix (1DM) \cite{coleman}, one can state that all Hamiltonians expressed as first-order operators of the $g_\alpha$'s will be correctly optimized if the following `Gutzwiller' matrix positivity conditions are satisfied:
\begin{eqnarray}
\rho^G \succeq 0 \qquad\text{with}\qquad \rho^G_{\alpha\beta} = \bra{\Psi^N}g^\dagger_{\alpha}g_{\beta}\ket{\Psi^N}~,\\
q^G \succeq 0 \qquad\text{with}\qquad q^G_{\alpha\beta} = \bra{\Psi^N}g_{\alpha}g^\dagger_{\beta}\ket{\Psi^N}~.
\end{eqnarray}
These matrices can be expressed as a function of regular fermionic creation and annihilation operators,
\begin{eqnarray}
\label{gutz_rho}\rho^{G}_{\alpha\beta} &=& \rho_{\alpha\beta} - \Gamma_{\alpha\bar{\beta};\beta\bar{\beta}} - \Gamma_{\alpha\bar{\alpha};\beta\bar{\alpha}} + \bra{\Psi^N}a^\dagger_\alpha a^\dagger_{\bar{\alpha}} a_{\bar{\alpha}} a^\dagger_{\bar{\beta}} a_{\bar{\beta}} a_{\beta}\ket{\Psi^N}~,\\
\label{gutz_q}q^G_{\alpha\beta} &=& \delta_{\alpha\beta}\left(1 - \rho_{\alpha\alpha} - \rho_{\beta\beta}\right) + \Gamma_{\bar{\beta}\beta;\bar{\beta}\alpha} + \Gamma_{\bar{\alpha}\beta;\bar{\alpha}\alpha}+ \bra{\Psi^N}a^\dagger_{\bar{\alpha}} a_{\bar{\alpha}} a_{\alpha} a^\dagger_{\beta}a^\dagger_{\bar{\beta}} a_{\bar{\beta}}\ket{\Psi^N}~,
\end{eqnarray}
in which $\rho$ is the 1DM:
\begin{equation}
\rho_{\alpha\beta} = \bra{\Psi^N}a^\dagger_{\alpha}a_\beta\ket{\Psi^N}~.
\end{equation}
It is clear from Eqs. (\ref{gutz_rho}),(\ref{gutz_q}) that the 3DM plays an essential role in describing the strong correlation limit (which is why the two-index conditions fail) but one also sees that not the full three-body space is needed. In fact, the 2.5DM, defined as:
\begin{equation}
W^l|^{S(S_{ab};S_{cd})}_{ab;cd} = \frac{1}{2\mathcal{S} + 1}\sum_{\mathcal{M}}\bra{\Psi^{\mathcal{SM}}}{B^\dagger}|^{S(S_{ab})}_{abl} B|^{S(S_{cd})}_{cdl}\ket{\Psi^{\mathcal{SM}}}~,
\label{2.5DM}
\end{equation}
is the minimal object from which both the $PQG$ conditions and the Gutzwiller conditions can be derived, and for which basic matrix manipulations are still on two-particle space.  In Eq.~(\ref{2.5DM}) $B^\dagger$ creates three particles:
\begin{equation}
{B^\dagger}|^{S(S_{ab})}_{abl} =\left(\left[a^\dagger_a \otimes a^\dagger_b\right]^{S_{ab}}\otimes a^\dagger_l\right)^S_{S_z}~,
\end{equation}
on lattice sites $a,b$ and $l$, coupled to total spin $S$, spin projection $S_z$ and intermediary spin $S_{ab}$. Note that the spin-averaged ensemble is used \cite{atomic} for describing the $N$-particle state with total spin $\mathcal{S}$. Here one considers an equal weight ensemble of all spin projections $\mathcal{M}$, and as a result the 2.5DM has no $S_z$ dependence. The 2.5DM is a block-diagonal object, in the sense that it is the 3DM diagonal in one pair of spatial indices. It can be used as the central object in a variational approach, applying constraints that include the $PQG$ and Gutzwiller conditions. This is a generalization of an approach used by Mazziotti {\it et al.} \cite{mazziotti,hammond} in a discussion of partial three-positivity constraints. They used a 3DM diagonal in both spatial \emph{and} spin indices as a variational object in a study of the Lipkin spin model. Letting the spin index be off-diagonal allows us to construct a spin-coupled version of the 2.5DM, which leads to an increase in speed of the optimization (see {\it e.g.} \cite{atomic}). Apart from this, the increased flexibility of the 2.5DM is important as it captures more correlation, which leads to a better result for the ground-state energy. We find that going from the spin-uncoupled to the spin-coupled form removes about 20\% from the remaining discrepancy with the exact result, in the intermediate $U/t$ region

The first non-trivial constraint we impose on the 2.5DM is a consistency condition that ensures symmetry between the diagonal third index and the other indices. As an example, one of these relations reads:
\begin{eqnarray}
\label{lifting}
W^l|^{S(S_{ab}S_{cd})}_{ab;cb} &=& [S_{ab}][S_{cd}]\sum_{S_{al}S_{cl}}[S_{al}][S_{cl}] \left\{\begin{matrix}S & \frac{1}{2} & S_{al}\\ \frac{1}{2} & \frac{1}{2} & S_{ab}\end{matrix}\right\}\left\{\begin{matrix}S & \frac{1}{2} & S_{cl}\\ \frac{1}{2} & \frac{1}{2} & S_{cd}\end{matrix}\right\}W^b|^{S(S_{al}S_{cl})}_{al;cl}~,
\end{eqnarray}
with $[S] = \sqrt{2S+1}$. In addition, we add constraints that are analogous to the standard two- and three-index conditions in that they can be expressed as matrix maps of the 2.5DM that have to be positive semidefinite. The first condition is simply that the different blocks of $W$ have to be positive semidefinite:
\begin{equation}
W^l \succeq 0~.
\end{equation}
The other five conditions are spin-adapted generalizations of the lifting conditions introduced in \cite{mazziotti,hammond,mazz_book}, and are of the form:
\begin{equation}
\mathcal{L}\left(W\right)^l\succeq 0\qquad\text{with}\qquad
\mathcal{L}\left(W\right)^l|^{S(S_{ab};S_{cd})}_{ab;cd}=\frac{1}{2\mathcal{S}+1}\sum_{\mathcal{M}}\bra{\Psi^{\mathcal{SM}}}{B^\dagger}|^{S(S_{ab})}_{abl} B|^{S(S_{cd})}_{cdl}\ket{\Psi^{\mathcal{SM}}}~,
\end{equation}
in which the $B^\dagger$ consist of different combinations of creation and annihilation operators.  As an example of such a condition, consider $B^\dagger$ defined as:
\begin{equation}
{B^\dagger}^{S(S_{ab})}_{abl} = \left(\tilde{a}_a \otimes \left[a^\dagger_b\otimes a^\dagger_l\right]^{S_{ab}}_{M_{ab}}\right)^S~ \qquad\text{where}\qquad 
\tilde{a}_{am_a} = (-1)^{\frac{1}{2}+m_a}a_{a -m_a}~.
\label{G1}
\end{equation}
The various $\mathcal{L}$'s arise by considering $(aaa),(a^\dagger a^\dagger a),(aaa^\dagger),(aa^\dagger a^\dagger) \text{ and } (a^\dagger a a)$ combinations, and can all be expressed as a function of $W$ through the use of anticommutation relations and spin recoupling. 

%Here the 2.5DM conditions explenation is over, so numerics and results!
The numerical optimization of the 2.5DM under these positivity constraints is a semidefinite program, and exactly the same methods used for the optimization of the 2DM can be used \cite{maz_bp,nakata_last,primal_dual}. The scaling of the basic matrix manipulations in this optimization is $M^7$, as opposed to the full three-index conditions, which scale as $M^9$, with $M$ the size of single-particle Hilbert space.
\begin{figure}
\centering
\includegraphics[scale=0.7]{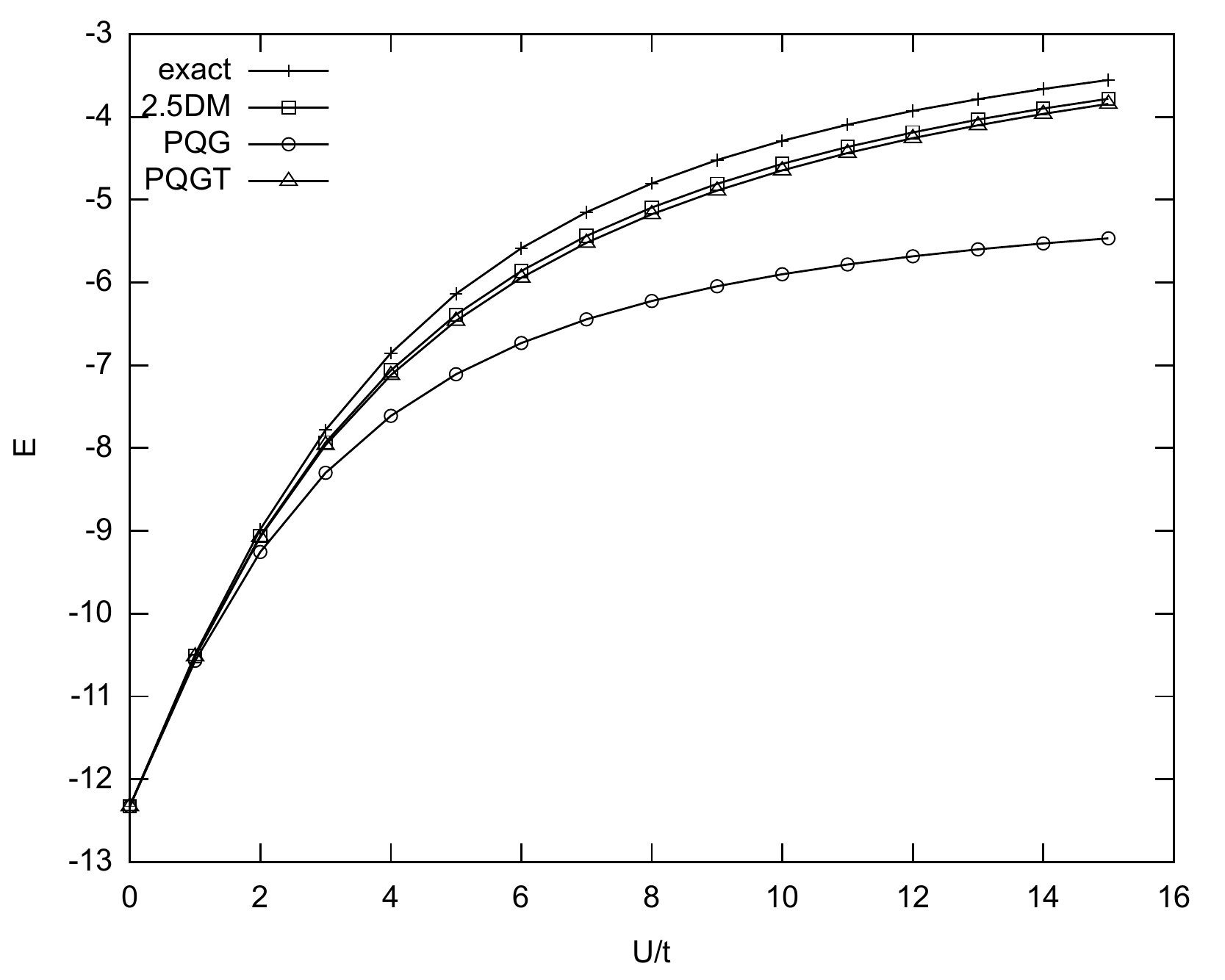}
\caption{\label{10_9} Ground-state energy as a function of on-site repulsion $U$ for 9 particles on a 10-site lattice. Exact results compared with v2.5DM results using 2.5-index conditions and v2DM results using $PQG$.}
\end{figure}
The result of such a v2.5DM calculation is also shown in Fig.~\ref{6_5}. It is clear that, as anticipated, the strong interaction limit is described with $PQGT$ quality without resorting to the full $PQGT$ framework. In fact, the v2.5DM results are slightly better then the $PQGT$. This is because the $T_1$ and $T_2$ conditions express the positivity of an anticommutator of three-particle operators, whereas in v2.5DM positivity is imposed on all possible individual products of three-particle operators, be it of a restricted class.
Similar results are obtained for a somewhat larger lattice of 10 sites, shown in Fig.~\ref{10_9}. In this case direct diagonalization is no longer an option, but we can compare to quasi-exact results calculated with a matrix product state (MPS) optimization \cite{verstraete,chan,sebastian}, which is uniquely suited for this kind of one-dimensional lattice problem. Again, the v2DM result using $PQG$ conditions is inaccurate, and one has to incorporate the three-particle correlations captured in the 2.5DM approach. The full-blown $PQGT$ calculation is far more costly than v2.5DM but produces slightly inferior results. Clearly both the $U\rightarrow 0$ and strongly interacting $U\rightarrow\infty$ limit are now exact. The latter statement is demonstrated in Tables \ref{tabel_5_6} and \ref{tabel_9_10} and follows from the above discussion about the Gutzwiller conditions. 
\begin{table}
\centering
\begin{tabular}{c|cccc}
$U$ & $PQG$ & $PQGT$ & v2.5DM&exact\\
\hline
50 & -4.89 & -2.54 &-2.53 & -2.46 \\
100 & -4.77  & -2.27 &-2.26 & -2.22\\
1000 & -4.67 & -2.03 & -2.03 &-2.02
\end{tabular}
\caption{ \label{tabel_9_10} Ground-state energy of a 10-site lattice with 9 particles for $U = 50,\ 100$ and 1000, exact results compared with v2DM results using $PQG$ and $PQGT$ results, and v2.5DM results.}
\end{table}

%conclusies en uitleg over veralgemening bij molecules enzoverder, 2D hub met symmetrie, prospects!
In summary, we developed the v2.5DM method that takes into account the necessary correlations needed to describe the large-$U$ limit of the Hubbard model, without having to resort to full-blown three-index conditions. It must be stressed that up to know we have only included the spin symmetry of the model in our code. If translational symmetry, parity and pseudospin symmetry are taken into account much larger lattices can be considered. As an example, our fully symmetric $PQG$ version allows lattice sizes up to 100 sites, and the fully symmetric $PQGT$ program up to 20 sites. We expect a fully symmetric version of v2.5DM to be applicable to lattice sizes of about 50 sites, thereby enabling us to study two-dimensional lattices of reasonable size.

The diagonality of the third index in the 2.5DM implies that the result will depend on the chosen single-particle basis. For the Hubbard model it is clear that the site basis is the optimal basis to use for the diagonal third index. It will be interesting to study other systems where it is less clear what the best choice of the single-particle basis would be. An appealing application, {\it e.g.}, are molecules, where one can hope to get three-index ($T_1T_2$) precision by applying v2.5DM method with a carefully chosen basis. A first guess of what the best basis would be is the basis of natural orbitals, for which it has been shown that the full-CI expansion has the fastest convergence \cite{lowdin}. This should be implemented using a outer self-consistency loop as a natural basis changes after each 2.5DM variational calculation.
\bibliography{gutzwiller.bib}
\end{document}